\documentclass[twocolumn,amsmath,amssymb,prd]{revtex4}

\def\al{\alpha}

\def\de{\delta}

\def\et{\eta}

\def\ka{\kappa}
\def\la{\lambda}

\def\rh{\rho}

\def\si{\sigma}

\def\ta{\tau}

\def\ph{\phi}

\def\La{\Lambda}

\def\cE{{\cal E}}

\def\cL{{\mathcal L}}

\def\fr#1#2{{{#1} \over {#2}}}
\def\half{{\textstyle{1\over 2}}}

\def\frac#1#2{{\textstyle{{#1}\over {#2}}}}

\def\vev#1{\langle {#1}\rangle}

\def\lsim{\mathrel{\rlap{\lower4pt\hbox{\hskip1pt$\sim$}}
    \raise1pt\hbox{$<$}}}
\def\gsim{\mathrel{\rlap{\lower4pt\hbox{\hskip1pt$\sim$}}
    \raise1pt\hbox{$>$}}}
\def\sqr#1#2{{\vcenter{\vbox{\hrule height.#2pt
         \hbox{\vrule width.#2pt height#1pt \kern#1pt
         \vrule width.#2pt}
         \hrule height.#2pt}}}}
\def\square{\mathchoice\sqr66\sqr66\sqr{2.1}3\sqr{1.5}3}

\newcommand{\beq}{\begin{equation}}
\newcommand{\eeq}{\end{equation}}
\newcommand{\bea}{\begin{eqnarray}}
\newcommand{\eea}{\end{eqnarray}}
\newcommand{\bit}{\begin{itemize}}
\newcommand{\eit}{\end{itemize}}
\newcommand{\rf}[1]{(\ref{#1})}

\begin{document}

\title{Constraints and Stability in Vector Theories with Spontaneous  Lorentz Violation}

\author{Robert Bluhm$^1$, Nolan L.\ Gagne$^1$, 
Robertus Potting$^2$, and Arturs Vrublevskis$^{1,3}$}

\affiliation{
$^1$Physics Department, Colby College,
Waterville, ME 04901 
\\
$^2$CENTRA, Departamento de F\'isica, 
Faculdade de Ci{\^ e}ncias e Tecnologia, Universidade do Algarve, Faro, Portugal
\\
$^3$Physics Department, 
Massachusetts Institute of Technology, 
Cambridge, MA 02139
}


\begin{abstract}
Vector theories with spontaneous Lorentz violation, 
known as bumblebee models, 
are examined in flat spacetime 
using a Hamiltonian constraint analysis.
In some of these models,
Nambu-Goldstone modes appear with
properties similar to photons in electromagnetism.
However,
depending on the form of the theory,
additional modes and constraints can appear
that have no counterparts in electromagnetism.
An examination of these constraints and
additional degrees of freedom,
including their nonlinear effects, 
is made for a variety of models with 
different kinetic and potential terms,
and the results are compared with electromagnetism.
The Hamiltonian constraint analysis also permits
an investigation of the stability of these models.
For certain bumblebee theories with a timelike vector,
suitable restrictions of the initial-value solutions are identified
that yield ghost-free models with a positive Hamiltonian.
In each case,
the restricted phase space is found to match that 
of electromagnetism in a nonlinear gauge.
\end{abstract}


\maketitle

\section{Introduction}

Investigations of quantum-gravity theories have uncovered
a variety of possible mechanisms that can lead to Lorentz violation.
Of these,
the idea that Lorentz symmetry might be spontaneously broken
\cite{ks}
is one of the more elegant.
Spontaneous Lorentz violation occurs when a vector or tensor
field acquires a nonzero vacuum expectation value.
The presence of these background values 
provides signatures of Lorentz violation that can 
be probed experimentally.
The theoretical framework for their investigation is given
by the Standard-Model Extension (SME)
\cite{sme,rbsme}.
Experimental searches for low-energy signals of Lorentz violation have
opened up a promising avenue of research in investigations of
quantum-gravity phenomenology
\cite{aknr,cpt}.

Theories with spontaneous Lorentz violation can also exhibit a variety
of physical effects due to the appearance of both Nambu-Goldstone 
(NG) and massive Higgs modes
\cite{rbak,rbffak,akrp}.
In the context of a gravitational theory,
these effects include modifications of gravitational propagation,
as well as altered forms of the static Newtonian potential,
both of which may be of interest in theoretical investigations 
of dark energy and dark matter.
Many investigations to date have concentrated on the case 
of a vector field acquiring a nonzero vacuum value.
These theories, called bumblebee models
\cite{ks,bb1,BBnote},
are the simplest examples of field theories with
spontaneous Lorentz breaking.
Bumblebee models can be defined 
with different forms of the potential and kinetic terms for the vector field, 
and with different couplings to matter and gravity 
\cite{baak,tj08,kt,jwm,bmg,cl,bp,chk}.
They can be considered as well in different spacetime geometries,
including Riemann, Riemann-Cartan, or Minkowski spacetimes.

Much of the interest in bumblebee models stems from the fact
that they are theories without local $U(1)$ gauge symmetry,
but which nonetheless allow for the propagation of massless vector modes.
Indeed,
one idea is that bumblebee models, 
with appropriate kinetic and potential terms, 
might provide alternative descriptions of photons 
besides that given by local $U(1)$ gauge theory.
In this scenario, 
massless photon modes arise as NG modes
when Lorentz violation is spontaneously broken.
However,
in addition to lacking local U(1) gauge invariance,
bumblebee models differ from electromagnetism
(in flat or curved spacetime) in a number of other ways.
For example,
the kinetic terms need not have a Maxwell form.
Instead, a generalized form as considered,
for example, 
in vector-tensor theories of gravity can be used,
though typically this may involve the introduction of
ghost modes into the theory.
Further differences arise due to the presence of a
potential term $V$ in the Lagrangian density for bumblebee models.
It is this term that induces spontaneous Lorentz breaking.
It can take a variety of forms,
which may involve additional excitations due to the presence
of massive modes or Lagrange-multiplier fields
that have no counterparts in electromagnetism.

The goal of this paper is to investigate further the extent to which
bumblebee models can be considered as equivalent to 
electromagnetism or as containing electromagnetism as a subset theory.
This question is examined here in flat spacetime.
While gravitational effects are a feature of 
primary interest in bumblebee models,
any equivalence or match to electrodynamics would presumably
hold as well in an appropriate flat-spacetime limit.
In a Minkowski spacetime,
the main differences between bumblebee models and electromagnetism
are due to the nature of the constraints imposed on the field variables 
and in the number of physical degrees of freedom permitted 
by the theory.
To investigate these quantities,
a Hamiltonian constraint analysis 
\cite{Dirac64,Hanson76,Gitman91,Henneaux}
is used.
This approach is particularly well suited for identifying the
physical degrees of freedom in a theory with constraints.
It can be carried out exactly with all nonlinear terms included.
It also permits examination of the question of whether the Hamiltonian 
is bounded from below over the constrained phase space.

\section{Bumblebee Models and Electromagnetism}

Bumblebee models are field theories with spontaneous Lorentz violation
in which a vector field acquires a nonzero vacuum value.
For the case of a bumblebee field $B_\mu$
coupled to gravity and matter,
with generalized quadratic kinetic terms 
involving up to second-order derivatives in $B_\mu$,
and with an Einstein-Hilbert term for the pure-gravity sector,
the Lagrangian density is given as
\bea
\cL_B &=& \fr 1 {16 \pi G} (R - 2 \La)
+ \si_1 B^\mu B^\nu R_{\mu\nu}
+ \si_2 B^\mu B_\mu R
\nonumber \\
&& - \fr 1 4 \ta_1 B_{\mu\nu} B^{\mu\nu} 
+ \fr 1 2 \ta_2 D_\mu B_\nu D^\mu B^\nu
\nonumber \\
&&+ \fr 1 2 \tau_3 D_\mu B^\mu D_\nu B^\nu  
- V(B_\mu B^\mu \mp b^2) + \cL_{\rm M} .
\label{LB}
\eea
In this expression, 
$b^2 >0$ is a constant,
and in Riemann spacetime
$B_{\mu\nu} = \partial_\mu B_\nu - \partial_\nu B_\mu$.
The quantities $\si_1$, $\si_2$, $\ta_1$, $\ta_2$, and $\ta_3$ 
are fixed constants that determine the form of the 
kinetic terms for the bumblebee field.
The term $\cL_{\rm M}$ represents possible interaction terms
with matter fields or external currents.
The potential $V(B_\mu B^\mu \mp b^2)$ has a minimum with respect to
its argument or is constrained to zero when
\beq
B_\mu B^\mu \mp b^2 = 0 .
\label{condition}
\eeq
This condition is satisfied when the vector field has a
nonzero vacuum value
\beq
B_\mu = \vev{B_\mu} = b_\mu ,
\label{vev}
\eeq
with  $b_\mu b^\mu = \pm b^2$.
It is this vacuum value that spontaneously breaks Lorentz invariance.

There are many forms that can be considered for 
the potential $V(B_\mu B^\mu \mp b^2)$.
These include functionals involving Lagrange-multiplier fields,
as well as both polynomial and nonpolynomial functionals
in $(B_\mu B^\mu \mp b^2)$
\cite{ks,baak}.
In this work,
three limiting-case examples are considered.
They represent the dominant leading-order terms that would
arise in an expansion of a general scalar potential $V$,
comprised of vector fields $B_\mu$,
which are not simply mass terms.
They include examples that are widely used in the literature.
The first introduces a Lagrange-multiplier field $\la$
and has a linear form,
\beq
V = \la (B_\mu B^\mu \mp b^2) .
\label{Vsigma}
\eeq
which leads to the constraint \rf{condition}
appearing as an equation of motion.
The second is a smooth quadratic potential
\beq
V = \half \ka (B_\mu B^\mu \mp b^2)^2 ,
\label{Vkappa}
\eeq
where $\ka$ is a constant.
The third again involves a Lagrange-multiplier field $\la$,
but has a quadratic form,
\beq
V = \half \la (B_\mu B^\mu \mp b^2)^2 . 
\label{Vsigma2}
\eeq
With this form,
the Lagrange multiplier field $\la$ decouples from the
equations of motion for $B_\mu$.

The model given in \rf{LB} involving a vacuum-valued vector 
has a number of features considered previously in the literature.
For example, 
with the potential $V$ and the cosmological constant $\La$ excluded,
the resulting model has the form of a vector-tensor theory of gravity 
considered by Will and Nordvedt
\cite{wn,cmw}.
Models with potentials \rf{Vsigma} and \rf{Vkappa} inducing 
spontaneous symmetry breaking
were investigated by Kosteleck\'y and Samuel (KS) 
\cite{ks},
while the potential \rf{Vsigma2} was recently examined in
\cite{rbffak}.
The special cases with a nonzero potential $V$,
$\ta_1 =1$, and $\si_1 = \si_2 = \ta_2 = \ta_3 = 0$
are the original KS bumblebee models
\cite{ks}.
Models with a linear Lagrange-multiplier potential \rf{Vsigma},
$\si_1 = \si_2 = 0$,
but arbitrary coefficients $\ta_1$, $\ta_2$, and $\ta_3$ 
are special cases 
(with a fourth-order term in $B_\mu$ omitted) 
of the models described in Ref.\ \cite{tj08}.

Since bumblebee models spontaneously break 
Lorentz and diffeomorphism symmetry,
it is expected that massless Nambu-Goldstone (NG)
and massive Higgs modes should appear in these theories.
The fate of these modes was recently investigated in
\cite{rbak,rbffak}.
The example of a KS bumblebee was considered in detail.
It was found that for all three potentials
\rf{Vsigma}, \rf{Vkappa}, and \rf{Vsigma2},
massless NG modes can propagate and
behave essentially as photons.
However,
in addition,
it was found that massive modes can appear 
that act as additional sources of energy and charge density.
In a linearized and static limit of the KS bumblebee,
it was shown that both the Newtonian and Coulomb potentials 
for a point particle are altered by the presence of a massive mode.
Nonetheless,
with suitable choices of initial values, 
which limit the phase space of the theory, 
solutions equivalent to those in Einstein-Maxwell theory 
can be obtained for the KS bumblebee models.

Bumblebee models with other (non-Maxwell) values of 
the coefficients $\ta_1$, $\ta_2$, and $\ta_3$ are expected
to contain massless NG modes as well.
However, in this case,
since the kinetic terms are different,
a match with electrodynamics is not expected.
The non-Maxwell kinetic terms alter
the constraint structure of the theory significantly, 
and a different number of physical degrees 
of freedom can emerge.

To compare the constraint structures of
different types of bumblebee models
with each other and with electrodynamics,
the flat-spacetime limit of \rf{LB} is considered.
The Lagragian density in this case reduces to 
\bea
\cL &=& - \fr 1 4 \ta_1 B_{\mu\nu} B^{\mu\nu} 
+ \fr 1 2 \ta_2 \partial_\mu B_\nu \partial^\mu B^\nu 
\nonumber \\
&&+ \fr 1 2  \ta_3 \partial_\mu B^\mu \partial_\nu B^\nu
- V(B_\mu B^\mu \pm b^2) - B_\mu J^\mu .
\label{L}
\eea
For simplicity, 
interactions consisting of
couplings with an externally prescribed current $J^\mu$
are assumed,
and a  Minkowski metric $\et_{\mu\nu}$ in Cartesian
coordinates with signature $(+,-,-,-)$ is used.

Following a Lagrangian approach,
second-order differential equations of motion for $B^\mu$ are obtained.
They are:
\bea
&&(\ta_1 + \ta_3) \left[  \square B_\mu - \partial_\mu \partial^\nu B_\nu \right]
\nonumber \\
&& \quad\quad\quad\quad
- (\ta_2 + \ta_3) \square B_\mu - 2 V^\prime B_\mu - J_\mu = 0 .
\label{BBeqs}
\eea
Here,
$V^\prime$ denotes variation of the potential $V(X)$ with 
respect to its argument $X$.
Since the NG modes stay in the minimum of the potential,
a nonzero value of $V^\prime$ indicates the presence of 
a massive-mode excitation.
Taking the divergence of these equations gives
\beq
\partial^\mu \left[ (\ta_2 + \ta_3)  \square B_\mu
+ 2 V^\prime B_\mu + J_\mu  \right] = 0 .
\label{dBBeqs}
\eeq
Clearly, as expected,
with $V=V^\prime=0$, $\ta_1 = 1$, 
and the remaining coefficients set to zero,
the equations of motion reduce to those of electrodynamics,
and \rf{dBBeqs} reduces to the statement of current conservation.
However,
if a nonzero potential with $V^\prime \ne 0$,
or if arbitrary values of $\ta_1$, $\ta_2$, $ \ta_3$ are allowed,
then a modified set of equations holds.

In flat spacetime,
the KS bumblebee has a nonzero potential $V$
and coefficients $\ta_1 =1$, and $\ta_2 = \ta_3 = 0$.
Its equations of motion evidently have a close resemblance 
to those of electrodynamics.
The main difference is that the KS bumblebee field itself acts
nonlinearly as a source of current.
Equation \rf{dBBeqs} shows that the matter current $J_\mu$
combines with the term $2 V^\prime B_\mu$ to form
a conserved current.

Interestingly, if the matter current $J_\mu$ is set to zero,
and a linear Lagrange-multiplier potential \rf{Vsigma} is used,
the KS model in flat spacetime reduces to a theory considered by Dirac
long before the notion of spontaneous symmetry breaking 
had been introduced
\cite{dirac51}.
Dirac investigated a vector theory 
with a nonlinear constraint identical to \rf{condition} 
with the idea of finding an alternative explanation of electric charge.
In his model, 
gauge invariance is destroyed,
and conserved charge currents appear only as 
a result of the nonlinear term involving $V^\prime$
for the Lagrange-multiplier potential.
Dirac did not,
however, 
propose a theory of Lorentz violation.
A vacuum value $b_\mu$ was never introduced,
and with $J_\mu = 0$ no Lorentz-violating interactions
with matter enter in the theory.

The idea that the photon could emerge as NG modes
in a theory with spontaneous Lorentz violation 
came more than ten years after the work of Dirac.
First, Bjorken proposed a model in which collective excitations
of a fermion field could lead to composite photons emerging as NG modes
\cite{bjorken}.
The observable behavior of the photon in this original model was 
claimed to be equivalent to electrodynamics.
Subsequently,
Nambu recognized that the constraint
\rf{condition} imposed on a vector field could also lead
to the appearance of NG modes that behave like photons
\cite{nambu68}.
He introduced a vector model 
that did not involve a symmetry-breaking potential $V$.
Instead, the constraint \rf{condition} was imposed as a
nonlinear U(1) gauge-fixing condition 
directly at the level of the Lagrangian.
The resulting gauge-fixed theory thus contained only three 
independent vector-field components in the Lagrangian.
Nambu demonstrated that his model was equivalent to 
electromagnetism and stated that the vacuum vector can be 
allowed to vanish to restore full Lorentz invariance.

In contrast to these early models,
the KS bumblebee was proposed as a theory with
physical Lorentz violation.
Even if the NG modes are interpreted as photons
in the KS model,
and no massive modes are present,
interactions between the vacuum vector $b_\mu$ and the
matter current $J_\mu$ provide clear 
observable signals of physical Lorentz violation.
However,
the presence of a potential $V$ also allows
additional degrees of freedom to enter in the KS model.
If arbitrary values of the coefficients $\ta_1$, $\ta_2$, and $\ta_3$
are permitted as well,
the resulting theory can differ substantially from electromagnetism.

Since many of these models contain unphysical modes,
either as auxiliary or Lagrange-multiplier fields,
constraint equations are expected to hold.
It is the nature of these constraints that determines ultimately
how many physical degrees of freedom occur in a given model.
With Dirac's Hamiltonian constraint analysis,
a direct procedure exists for determining the constraint structure 
and the number of physical degrees of freedom in these models.

\section{Hamiltonian Constraint Analysis}

Given a Lagrangian density $\cL$
describing a vector field $B_\mu$,
the canonical Hamiltonian density is 
${\cal H} = \Pi^\mu\partial_0 B_\mu - \cL$,
where the canonical momenta are defined as
\beq
\Pi^\mu = \fr{\de \cL}{\de(\partial_0 B_\mu) } .
\label{Pi}	 
\eeq
If additional fields, e.g., Lagrange multipliers $\la$,
are contained in the theory,
additional canonical momenta for these quantities
are defined as well,
e.g., $\Pi^{(\la)}$.
(Note: here $\la$ is not a spacetime index).
In the Hamiltonian approach,
time derivatives of a quantity $f$ are computed by
taking the Poisson bracket with the Hamiltonian $H$,
\beq
\dot f = \{f, H \} + \fr{\partial f}{\partial t} .
\label{fdot}
\eeq
The second term is needed with quantities
that have explicit time dependence,
e.g., an external current $J^\mu$.

In Dirac's constraint analysis,
primary and secondary constraints are determined,
and these are identified as either first-class or second-class.
In the phase space away from the constraint surface,
the canonical Hamiltonian is ambiguous 
up to additional multiples of the constraints.
An extended Hamiltonian is formed that includes multiples
of the constraints with coefficients that can be determined,
or in the case of first-class constraints,
remain arbitrary.
It is the extended Hamiltonian that is then used in \rf{fdot}
to determine the equations of motion for the fields
and conjugate momenta.

A system of constraints is said to be regular 
if the Jacobian matrix formed from variations of the constraints 
with respect to the set of field variables and conjugate 
momenta has maximal rank.
If it does not,
the system is said to be irregular,
and some of the constraints are typically redundant.
Dirac argued that theories with primary first-class constraints
have arbitrary or unphysical degrees of freedom,
such as gauge degrees of freedom.
These types of constraints therefore allow removal of
two field or momentum components.
Dirac conjectured that this is true as well for secondary
first-class constraints.
Based on this,
a counting argument can be made.
It states that in a theory with $n$ field 
and $n$ conjugate-momentum components,
if there are $n_1$ first-class constraints and $n_2$
second-class constraints,
the number of physical independent degrees of freedom
is $n - n_1 - n_2/2$.
(Note: it can be shown that $n_2$ is even).
This counting argument based on Dirac's conjecture
holds up well for theories with regular systems of constraints.
However,
counterexamples are known for irregular systems
\cite{Henneaux}.

Once the unphysical modes have been eliminated,
by applying the constraints and/or imposing gauge conditions,
the evolution of a physical system is determined by the 
equations of motion for the physical fields and momenta,
subject to initial conditions for these quantities.
Any bumblebee theory that has additional degrees of freedom in
comparison to electrodynamics must therefore specify additional
initial values.
The subsequent evolution of the extra degrees of freedom
typically leads to effects that do not occur in electrodynamics.
However, in some cases,
equivalence with electrodynamics can hold in a subspace
of the phase space of the modified theory.
For this to occur,
initial values must exist that confine the evolution of 
the theory to a region of phase space that matches 
electrodynamics in a particular choice of gauge.

In general,
the stability of a theory,
e.g., whether the Hamiltonian is positive,
depends on the initial values and allowed
evolution of the physical degrees of freedom.
As discussed in the subsequent sections,
most bumblebee models contain regions of phase space
that do not have a positive definite Hamiltonian,
though in some cases,
restricted subspaces can be found that do 
maintain ${\cal H}>0$.
In a quantum theory,
instability in any region of the classical phase space
might be expected to destabilize the full theory.
However,
bumblebee models,
with gravity included,
are intended as effective theories presumably emerging 
at or below the Planck scale from a more fundamental 
(and unknown) quantum theory of gravity.
In this context,
quantum-gravity effects might impose additional
constraints leading to stability.
However,
in the absence of a fundamental theory, 
the question of the ultimate stability of 
bumblebee models cannot be addressed.
For this reason, 
in the subsequent sections,
only the behavior of bumblebee models
in classical phase space is considered.

The following sections apply Dirac's constraint analysis to 
a number of different bumblebee models,
including the KS bumblebee as well as more general cases
with arbitrary values of the coefficients $\ta_1$, $\ta_2$, $\ta_3$.
Since much of the literature has focused on the case
of a timelike vector $B_\mu$,
this restriction is assumed throughout this work as well.
With this assumption,
there always exists an observer frame
in which rotational invariance is maintained and
only Lorentz boosts are spontaneously broken.
For each type of model to be considered, 
all three of the potentials in 
\rf{Vsigma}, \rf{Vkappa}, and \rf{Vsigma2} are considered.
For comparison (and use as benchmarks),
electromagnetism and the theory of Nambu
are considered as well.
In each case,
the explicit form of the Lagrangian is obtained from 
\rf{L}  by inserting appropriate values for 
$V$, $\ta_1$, $\ta_2$, and $\ta_3$,
and the conjugate momenta and Hamiltonian are then computed.
For example, electrodynamics is obtained by setting $V=0$,
$\ta_1=1$, and $\ta_2 = \ta_3 = 0$.
Conventional notation sets $B_\mu = A_\mu$
and $B_{\mu\nu} = F_{\mu\nu}$.
The Hamiltonian is given in terms of 
the four fields $A_\mu$ and their conjugate momenta $\Pi^\mu$.
The Lagrangian in Nambu's model also starts with these same
values (allowing U(1) invariance).
However, in this case,
one component of $A_\mu$ is eliminated in terms
of the remaining three, 
using the nonlinear
condition in \rf{condition}.
For the case of a timelike vector,
the substitution $A_0 = (b^2 + A_j^2)^{1/2}$
is made directly in $\cL$.
The resulting Hamiltonian in Nambu's model
therefore depends only on three fields
$A_j$ and three conjugate momenta $\Pi^j$.
In contrast,
bumblebee models are defined with a nonzero potential $V$
and have Hamiltonians that depend on all four fields $B_\mu$
and their corresponding conjugate momenta $\Pi^\mu$.
Examples with a Lagrange-multiplier potential involve
a fifth field $\la$ and its conjugate momentum $\Pi^{(\la)}$.
However, in examples with a smooth quadratic potential,
there is no Lagrange multiplier,
and the relevant fields and momenta
are $B_\mu$ and $\Pi^\mu$.

\subsection{Electromagnetism}

The conjugate momenta in electrodynamics are
\beq
\Pi^j = \partial_0 A_j - \partial_j A_0 ,
\quad\quad
\Pi^0 = 0 .
\eeq
The latter constitutes a primary constraint,
$\ph_1 = \Pi^0 \approx 0$.
It leads to a secondary constraint,
$\ph_2 = \partial_j \Pi^j - J^0 \approx 0$,
which is Gauss' law, 
since $\Pi^j$ can
be identified as the electric field components $E^j$
and $J^0$ is the charge density.
In these expressions and below, 
Dirac's weak equality symbol
``$\approx$'' is used to denote equality on the submanifold defined by
the constraints \cite{Henneaux}.
Both of the constraints $\ph_1$ and $\ph_2$ are first-class,
indicating that there are gauge or unphysical degrees of freedom.
Following Dirac's counting argument,
there should be $n-n_1-n_2/2 = 4 - 2 - 0 = 2$
independent physical degrees of freedom.
These are the two massless transverse photon modes.

The canonical Hamiltonian in electrodynamics is
\beq
{\cal H} = \frac{1}{2}(\Pi^j)^2 + \Pi^j\partial_j A_0 
+ \fr 1 2 (F_{jk})^2
+ A_\mu J^\mu .
\label{HEM}
\eeq
In the presence of a static charge distribution,
with $J^\mu  = (\rh,\vec J) = (\rh(\vec x),0)$,
no work is done by the external current,
and the Hamiltonian is positive definite.
To observe this,
integrate by parts
and use the constraint $\ph_2$ (Gauss' law)
to show that ${\cal H} = \frac{1}{2}(\Pi^i)^2 + \fr 1 2 (F_{jk})^2 \ge 0$.

The equations of motion for the fields $A_\mu$ and
momenta $\Pi^\mu$ obtained from the extended
Hamiltonian contain arbitrary functions
due to the existence of the first-class constraints.
These can be eliminated by imposing gauge-fixing conditions.
The evolution of the physical degrees of freedom,
subject to a given set of initial values, 
is then determined for all time.

\subsection{Nambu's Model}

The starting point for Nambu's model
\cite{nambu68} 
is the conventional Maxwell
Lagrangian with U(1) gauge invariance and a conserved current $J^\mu$.
For the case of a timelike vector $A_\mu$,
the condition $A_0 = (b^2 + A_j^2)^{1/2}$ 
is substituted directly into the Lagrangian
as a gauge-fixing condition.
The result is
\bea
\cL &=&  \; \frac{1}{2}(\partial_0 A_j)^2 
+ \frac{1}{2}(\partial_j (b^2 + A_k^2)^{1/2})^2 
- \frac{1}{2}(\partial_j A_k)^2 
\nonumber \\
&&+ \frac{1}{2}(\partial_j A_k)(\partial_k A_j)
- (\partial_j (b^2 + A_k^2)^{1/2})(\partial_0 A_j) 
\nonumber \\
&&- (b^2 + A_k^2)^{1/2} J^0 - A_j J^j  .
 \label{LNambu}
\eea
Nambu claimed that this theory is equivalent 
to electromagnetism in a nonlinear gauge.
He argued that a U(1) gauge transformation 
exists that transforms an electromagnetic field
in a standard gauge into the field
$A_\mu$ obeying the nonlinear gauge
condition $A_\mu A^\mu = b^2$.

The Hamiltonian in Nambu's model is 
\bea
{\cal H} &=& \frac{1}{2}(\Pi^j)^2 + \fr 1 2 (F_{jk})^2 
+ \Pi^j \partial_j (b^2 + A_k^2)^{1/2} 
\nonumber \\
&&+ (b^2 + A_k^2)^{1/2} J^0
+ A_j J^j .
\label{HNam}
\eea
It depends on three field components $A_j$ 
and their conjugate momenta
$\Pi^j = \partial_0 A_j - \partial_j (b^2 + A_k^2)^{1/2}$.
In this theory,
there are no constraints, 
and therefore application of Dirac's counting 
argument says that there are three physical degrees of freedom,
which is one more than in electromagnetism.
An extra degree of freedom arises
because gauge fixing at the level of the Lagrangian
causes Gauss' law,
$\partial_j \Pi^j - J^0=0$, 
to disappear as a constraint equation.
A smilar disappearance of Gauss' law is known to
occur in electrodynamics in temporal gauge 
(with $A_0 = 0$ substituted in the Lagrangian)
\cite{rj80}.
Indeed,
the linearized limit of Nambu's model with a timelike
vector field is electrodynamics in temporal gauge.

Observe that with with $\vec J = 0$
and using integration by parts, 
the Hamiltonian can be rewritten as
\beq
{\cal H} = \frac{1}{2}(\Pi^j)^2 + \fr 1 2 (F_{jk})^2 
- (\partial_j \Pi^j - J^0) (b^2 + A_k^2)   .
\label{HNam22}
\eeq
In the absence of a constraint enforcing Gauss' law,
${\cal H}$ need not be positive definite.
For example,
if the extra degree of freedom in $A_j$ 
causes large deviations from Gauss' law,
which are not forbidden by any constraint,
then negative values of ${\cal H}$ can occur.

However,
equivalence between Nambu's model and
electrodynamics can be established by restricting the
phase space in Nambu's theory.
To see that  this follows,
consider the equations of motion in
Nambu's model,
\beq
\dot A_j = \Pi^j + \partial_j (b^2 + A_k^2)^{1/2} ,
\label{n1}
\eeq
\bea
\dot \Pi^j &=& \partial^k \partial_k A^j - \partial^j \partial_k A^k 
- \partial_l \Pi^l \fr {A^j} {(b^2 + A_k^2)^{1/2}}
\nonumber \\
&& \quad\quad\quad
+ \fr {A^j J^0} {(b^2 + A_k^2)^{1/2}} - J^j .
\label{n2}
\eea
Taking the spatial divergence of \rf{n2} and using current
conservation yields the nonlinear relation
\beq
\partial_0 (\partial_j \Pi^j - J^0) =
- \partial_j \left[ (\partial_l \Pi^l - J^0) \fr {A^j} {(b^2 + A_k^2)^{1/2}} \right] .
\label{NamGauss1}
\eeq
This equation shows that if Gauss' law, 
$(\partial_j \Pi^j - J^0)=0$,
holds at $t=0$,
then $\partial_0 (\partial_j \Pi^j - J^0)=0$ as well at $t=0$.
Together these conditions and Eq.\ \rf{NamGauss1}
are sufficient to show that Gauss' law then holds for all time.
From this it follows that
${\cal H}$ is positive over
the restricted phase space,
which matches that of electrodynamics
in a nonlinear gauge.
Thus,
by restricting the phase space to solutions with initial
values obeying Gauss' law,
the equivalence of Nambu's model
with electromagnetism is restored.

\subsection{KS Bumblebee Model}

KS bumblebee models
\cite{ks} 
in flat spacetime have a Maxwell kinetic term
and a nonzero potential $V$.
The choice of a Maxwell form for the kinetic term 
is made to prevent propagation of the longitudinal mode
of $B_\mu$ as a ghost mode.
The KS Lagrangian is obtained from \rf{L} 
by setting $\ta_1 = 1$ and $\ta_2 = \ta_3 = 0$.
The constraint structures for models with each 
of the three potentials \rf{Vsigma} - \rf{Vsigma2}
are considered.
For definiteness,
the case of a  timelike vector $B_\mu$ is assumed.

\subsubsection{Linear Lagrange-Multiplier Potential}

With a linear Lagrange-multiplier potential \rf{Vsigma},
an additional field component $\la$ is introduced in
addition to the four fields $B_0$ and $B_j$.
The conjugate momenta are
\beq
\Pi^0 = \Pi^{(\la)} = 0 , \quad 
\Pi^i = \partial_0 B_i - \partial_i B_0 ,
\label{KSpi}
\eeq
and the canonical Hamiltonian is
\bea
{\cal H} &=& \frac{1}{2}(\Pi^i)^2 + \Pi^i\partial_i B_0 
+ \frac{1}{2}(\partial_i B_j)^2 
- \frac{1}{2}(\partial_j B_i)(\partial_i B_j) 
\nonumber \\
&&+ \lambda(B_0^2 - B_i^2 - b^2) + B_\mu J^\mu .
\label{KSHam1}
\eea
Four constraints are identified as
\begin{align}
	& \phi_1 = \Pi^0 \\ 
	& \phi_2 = \Pi^{(\la)} \\
	& \phi_3 = \partial_i \Pi^i - 2 \lambda B_0  - J^0 
	\label{KSgauss1}\\
	&\phi_4 = - \left( B_0^2 - B_j^2 - b^2 \right) .
\end{align}
The constraints $\phi_1$ and $\phi_2$ are primary,
while $\phi_3$ and $\phi_4$ are secondary.
All four are second-class.

Applying Dirac's algorithm to determine the number
of independent degrees of freedom gives
$n - n_1 - n_2/2 = 5 -2 -2/2 = 3$.
Hence,
there is an extra degree of freedom in the KS bumblebee
model in comparison to electrodynamics.
It arises due the presence of the extra field $\la$
and the changes in the types of constraints.
Unlike electromagnetism,
there are no first-class constraints in the KS bumblebee,
which reflects the lack of gauge invariance.
The constraint $\ph_3$ gives a modified form of Gauss' law
in which the combination $2 \lambda B_0$ 
acts as a source of charge density.
Since $V^\prime = \la$ in this example,
any excitation of the field $\la$ is away from the
potential minimum and therefore acts effectively as
a massive Higgs mode
\cite{rbffak}.
In curved spacetime,
such a mode can modify both the gravitational
and electromagnetic potentials of a point particle.
However,
here, in flat spacetime,
the presence of $\la$ leads only to modifications
of the Coulomb potential.

The Hamiltonian with $\vec J = 0$ reduces,
after using $\ph_3$, $\ph_4$, and integration by parts, to 
\begin{equation}
{\cal H} = \fr{1}{2}(\Pi^j)^2  
+ \fr{1}{2}(B_{jk})^2 
- 2 \la B_0^2 .
\label{KSHam2}
\end{equation}
The full phase space of the theory 
on the constraint surface includes regions
in which ${\cal H}$ is negative due to the presence 
of the additional degree of freedom.
For example,
consider the case with $J^0 = 0$ and initial values 
\cite{clay}
$B_j = \partial_j \ph (\vec x)$
and  $\Pi^j = -\partial_j (b^2 + (\partial_k \ph)^2)^{1/2}$ at $t=0$,
where $\ph (\vec x)$ is an arbitrary time-independent scalar.
These give $B_{jk} = 0$ and
$B_0 = (b^2 + (\partial_j \ph)^2)^{1/2}$ at $t=0$.
Inserting these initial values in \rf{KSHam2}
reduces the Hamiltonian to 
${\cal H} = - \frac{1}{2}(\Pi^j)^2$ at $t=0$.
The corresponding initial value for $\la$ is
\beq
\la = - \fr 1 2 (b^2 + (\partial_j \ph)^2)^{-1/2} 
\left[ \vec \nabla^2 (b^2 + (\partial_k \ph)^2)^{1/2}  \right] .
\label{lamt0}
\eeq
Evidently,
the Hamiltonian in the classical KS bumblebee model 
can be negative when 
nonzero values of $\la$ are allowed.

However, if initial values are chosen that restrict the phase
space to values with $\la = 0$,
the resulting solutions for the vector field
and conjugate momentum
are equivalent to those in electromagnetism
in a nonlinear gauge.
Examination of the equation of motion for $\la$,
\beq
\dot{\la} = \fr {1}{B_0 }\partial_j \left( \lambda B_j \right) 
- \fr {1}{2B_0}\partial_\mu J^\mu 
- \lambda \fr {B_j }{(B_0)^2}\left(\Pi^j  + \partial_j B_0 \right) ,
\label{ladot}
\eeq
reveals that if the current $J^\mu$ is conserved,
and $\la = 0$ at time zero, 
then $\la$ will remain zero for all time.
The Hamiltonian in this case is positive.
The equations of motion for $B_j$ and $\Pi^j$ are
\bea
\dot{B}_j &=& \Pi ^j  + \partial_j B_0 ,
\label{Bjeq}
\\
\dot{\Pi}^j &=& \partial _k \partial _k B_j  - \partial _j \partial _k B_k - J^j +2 \la B_j 
\label{Pijeq}
\eea
With $\la = 0$,
these combine to give the usual Maxwell equations
describing massless transverse photons.
The third component in $B_j$ is an auxiliary field
that is constrained by the usual form of Gauss' law
when $\la = 0$.
Note, however, that even with the phase space restricted to
regions with $\la = 0$,
the matter sector of the theory will exhibit signatures of
the spontaneous Lorentz violation through the interaction
of the vacuum value $b_\mu$ with the matter current $J^\mu$.

It is clear from these results,
that conservation of the matter current $J^\mu$ is
necessary for the stability of the KS bumblebee model.
Note, however, that the theories lack local U(1) gauge invariance
and that the current conservation could arise simply from
matter couplings that are invariant under a global U(1) symmetry.
As a result, photons in the KS bumblebee model appearing as NG modes
are due to spontaneous Lorentz breaking, 
not local U(1) gauge invariance.
For further discussion of the bumblebee currents,
including in the presence of gravity,
see Ref.\ \cite{rbffak}.
In that work,
there is also further discussion of the fact that 
the Lagrange-multiplier field can act as a source 
of charge density in the KS bumblebee model
and that there can exist solutions
(with nonzero values of $\la$)
in which the field lines converge or become singular,
even in the absence of matter charge.
This behavior has been referred to in the literature
as the formation of caustics in the KS model.
However,
as described in \cite{rbffak},
it is simply a natural consequence of the fact that the bumblebee 
fields themselves act as sources of current.
Moreover,
with the phase space restricted to regions with $\la = 0$,
the only singularities appearing 
for the case of a timelike vector $B_\mu$ 
are those due to the presence of matter charge 
as in ordinary electrodynamics with a $1/r$ potential.

\subsubsection{Quadratic Smooth Potential}

A similar analysis can be performed for a KS bumblebee with
the smooth quadratic potential defined in \rf{Vkappa}.
The parameter $\ka$ appearing in $V$ is a constant.
Therefore,
in this case,
there are four fields $B_0$, $B_j$,
and their four conjugate momenta,
\beq
\Pi^0 = 0 , \quad
\Pi^j = \partial_0 B_j - \partial_j B_0 .
\label{KSsmoothPi}
\eeq
There are two constraints,
\bea
\ph_1 &=& \Pi^0  \\
\phi_2 &=& \partial_j \Pi^j - 2\kappa B_0  \left( B_0^2 - B_j^2 - b^2 \right) - J^0 ,
\label{KSsmoothPhi}
\eea
where $\ph_1$ is primary, $\phi_2$ is secondary,
and both are second-class.
Dirac's counting argument says there are
$n - n_1 - n_2/2 = 4 - 0 - 2/2 = 3$ independent 
degrees of freedom,
which again is one more than in electromagnetism.

The condition \rf{condition} does not occur
as a constraint in this case.
Instead,
an extra degree of freedom appears
as a massive Higgs excitation
$V^\prime = 2\kappa B_0  \left( B_0^2 - B_j^2 - b^2 \right) \ne 0$
away from the potential minimum.
The constraint $\ph_2$ yields a modified version of Gauss' law,
showing that the massive mode acts as a source of
charge density.

The stability of the Hamiltonian with $\vec J = 0$ can be examined.
Using the constraints and integration by parts gives
\begin{equation}
\label{ka Hamiltonian}
{\cal H} = \fr{1}{2}(\Pi^j)^2 + \fr{1}{2}(B_{jk})^2 
- \fr{1}{2}\kappa (3 B_0^2 + B_j^2 + b^2) (B_0^2 - B_k^2 - b^2) ,
\end{equation}
which evidently is not positive over the full phase space.
If a nonzero massive mode proportional to $(B_0^2 - B_j^2 - b^2)$
is present,
negative values of ${\cal H}$ can occur.

However, equivalence to electrodynamics does hold in a restricted region
of phase space.
To verify this,
consider the equations of motion,
\bea
2 \ka \dot B_0 &=& (3B_0^2  - B_j^2 - b^2)^{-1}
\left[ 4 \ka B_0 B_k (\Pi^k + \partial_k B_0) \right.
\nonumber \\
&& \left.
+ 2 \ka \partial_k [B_k(B_0^2 - B_l^2 - b^2)] + \partial_\mu J^\mu \right] ,
\\
\label{KSsmooth1}
\dot B_j &=& \Pi^j + \partial_j B_0 ,
\label{KSsmooth2}
\\
\dot \Pi^0 &=& \partial_j \Pi^j - J^0 - 2 \ka B_0 (B_0^2 - B_j^2 - b^2) ,
\label{KSsmooth3}
\\
\dot \Pi^j &=& \partial_k \partial_k B_j - \partial_j \partial_k B_k
\nonumber \\
&&+ 2 \ka B_j (B_0^2 - B_k^2 - b^2) - J^j .
\label{KSsmooth4}
\eea
Combining these gives
\bea
\ka \partial_0 (B_0^2 - B_j^2 - b^2) =
(3B_0^2  - B_j^2 - b^2)^{-1} 
\quad\quad\quad\quad
\nonumber \\
\times \left[ 2 \ka B_0 \partial_k [B_k(B_0^2 - B_l^2 - b^2)]
+ B_0 \partial_\mu J^\mu
 \right.
\nonumber \\
\left. 
- 2 \ka (B_0^2 - B_k^2 - b^2)
B_l (\Pi^l + \partial_l B_0) 
 \right] .
\quad
\label{mmcond}
\eea
This equation reveals that if the current $J^\mu$ is
conserved and $(B_0^2 - B_j^2 - b^2) = 0$ at $t=0$,
then $(B_0^2 - B_j^2 - b^2) = 0$ for all time.
Therefore,
with these conditions imposed,
the massive mode never appears,
the Hamiltonian is positive,
and the phase space is restricted to solutions
in electromagnetism in the nonlinear gauge 
\rf{condition}.

In theories with a nonzero massive mode,
the size of the mass scale $\ka b^2$ becomes relevant.
For very large values,
perturbative excitations that go up the potential 
minimum would be expected to be suppressed.
Since the mass scale associated with spontaneous Lorentz 
violation is presumably the Planck scale,
its appearance necessarily brings gravity into the discussion.
It is at the Planck scale where quantum-gravity effects
might impose additional constraints that could maintain the
overall stability of the theory.
At sub-Planck energies,
massive-mode excitations have been shown to 
exert effects on classical gravity.
For example,
as shown in Ref.\ \cite{rbffak},
the gravitational potential of a point particle is modified.
However, in the limit where the mass of the massive mode 
becomes exceptionally large,
it was found for the case of the KS bumblebee model
that both the usual Newtonian and Coulomb 
potentials are recovered.

\subsubsection{Quadratic Lagrange-Multiplier Potential}

The KS bumblebee model with a quadratic Lagrange-multiplier
potential \rf{Vsigma2} involves five fields $\la$ and $B_\mu$.
In a Lagrangian approach,
the constraint \rf{condition} follows from the
equation of motion for $\la$.
The on-shell equations of motion for $B_\mu$ are 
the same as in electromagnetism.
In this case,
the field $\la$ decouples and does not act as a 
source of charge density.
On shell,
the potential obeys $V^\prime = 0$, 
current conservation $\partial_\mu J^\mu = 0$ holds, 
and there is no massive mode.
This model provides an example of a theory with physical
Lorentz violation due to the matter couplings with $J^\mu$.
Nonetheless,
in the electromagnetic sector,
the theory is equivalent to electromagnetism in the
nonlinear gauge \rf{condition}.

However,
the Hamiltonian formulation of this model involves an 
irregular system of constraints
\cite{Henneaux}.
Thus, depending on how the constraints are handled,
Dirac's counting algorithm might not apply
and equivalence with the Lagrangian approach may not hold.
The conjugate momenta are
\begin{align}
	& \Pi^0 = 0 \label{sq pi_0 v2} , \\
	& \Pi^j = \partial_0 B_j - \partial_j B_0 \label{sq pi_i} , \\
	& \Pi^{(\la)} = 0 .
	\label{sq pi_lambda v2}
\end{align}
From these, four constraints can be identified,
\begin{align}
	& \phi_1 = \Pi^0  \label{sq pi_0} , \\
	& \phi_2 = \Pi^{(\la)} \label{sq pi_lambda} , \\
	& \phi_3 = \partial_j \Pi^j 
- 2\lambda B_0  \left( B_0^2 - B_j^2 - b^2 \right) - J^0 , \\
	& \phi_4 = - \frac{1}{2}\left( B_0^2 - B_j^2 - b^2 \right)^2 .
\end{align}
With $\ph_4 \approx 0$,
the constraint surface is limited to fields obeying
$(B_0^2 - B_j^2 - b^2 ) = 0$,
and $\ph_3$ reduces to Gauss' law.
In this case,
$\ph_2$ and $\ph_4$ can be identified as first-class,
while $\ph_1$ and $\ph_3$ are second-class.
Dirac's counting algorithm then states that there are
$n - n_1 - n_2/2 = 5 - 2 - 2/2 = 2$ independent degrees
of freedom,
which matches electromagnetism,
and the Hamiltonian is positive throughout the full
physical phase space.
However,
if instead the squared constraint $\ph_4$ is replaced by
the equivalent constraint $\ph_4^\prime = (B_0^2 - B_j^2 - b^2)$
that spans the same constraint surface,
then a different set of results holds.
In this case,
additional constraints appear 
from the Poisson-bracket relations that are not equivalent
to the set defined above,
and Dirac's counting algorithm fails to determine
the correct number of degrees of freedom.
The resulting theory with $\ph_4^\prime$ replacing
$\ph_4$ is not equivalent to the Lagrangian approach.

Evidently,
care must be used in working with a squared constraint equation.
The constraints $\ph_4^\prime$ and $\ph_4$ are redundant,
and the Hamiltonian system is irregular.
Nonetheless,
with these caveats,
the KS model with a squared Lagrange-multiplier potential
provides a useful model of spontaneous Lorentz violation.
It allows an implementation of the symmetry breaking that
does not require enlarging the phase space to include a
massive mode or nonlinear couplings with $\la$.
The only physical degrees of freedom in the theory
are the NG modes that behave as photons.

\subsection{Bumblebee Models with $(\ta_2 + \ta_3) \ne 0$}

In this section,
the constraint analysis is applied to bumblebee models
in flat spacetime that have a Lagrangian \rf{L} with
a generalized kinetic term obeying $(\ta_2 + \ta_3) \ne 0$.
Such models do not have a Maxwell form for the kinetic term.
Throughout this section, 
arbitrary values of $\ta_1$, $\ta_2$, and $\ta_3$ 
are used;
however, 
it is assumed that discontinuities are avoided when these 
parameters appear in the denominators of equations.
The three potentials in \rf{Vsigma} - \rf{Vsigma2} are considered,
and $B_\mu$ is assumed to be timelike.
Since the kinetic term is not of the Maxwell form,
it is not expected that the NG modes in these types of models 
can be interpreted as photons.
For this reason,
the interaction term $B_\mu J^\mu$ is omitted in this section.

The point of view here is that the generalized bumblebee models
originate from a vector-tensor theory of gravity
with spontaneous Lorentz violation induced by the potential $V$.
In this context,
the vector fields $B_\mu$ have no matter couplings
and reduce to sterile fields in a flat-spacetime limit.
Nonetheless,
NG modes and massive modes can appear in this limit.
Dirac's Hamiltonian analysis is used to examine the
constraint structure and the number of physical degrees
of freedom associated with these modes.
Comparisons can then be made with the results in electromagnetism 
and the KS bumblebee models.

\subsubsection{Linear Lagrange-Multiplier Potential}

Beginning with a model with the linear Lagrange-multiplier 
potential in Eq.\ \rf{Vsigma},
the Lagrangian is given in terms of the five fields
$B_0$, B$_j$, and $\la$.
From this the conjugate momenta are found to be
\begin{align}
& \Pi^0 = (\ta_2 + \ta_3)(\partial_0 B_0) - \ta_3 (\partial_j B_j) , 
\label{Pi0lam} \\
& \Pi^j = (\ta_1 - \ta_2) (\partial_0 B_j) - \ta_1 (\partial_j B_0) , 
\label{Pijlam} \\
& \Pi^{(\la)} = 0
\label{Pilamlam} .
\end{align}
The canonical Hamiltonian is then given as
\bea
{\cal H} &=& 
\left(\fr{\ta_1^2 - (\ta_1 - \ta_2)^2}{2(\ta_1 - \ta_2)}\right)
(\partial_j B_0)^2 
+ \left(\fr{1}{2(\ta_1 - \ta_2)}\right)(\Pi^j)^2 
\nonumber \\
&&
+ \left(\fr{\ta_1}{\ta_1 - \ta_2}\right)\Pi^j (\partial_j B_0)
+ \fr 1 2 (\ta_1 - \ta_2) (\partial_j B_k)^2 
\nonumber \\
&& -\fr 1 2 \ta_1 (\partial_j B_k)(\partial_k B_j) 
+\left(\fr{1}{2(\ta_2 + \ta_3)}\right)(\Pi^0)^2 
\nonumber \\
&&
+\left(\fr{\ta_3}{\ta_2 + \ta_3}\right)\Pi^0 \partial_j B_j
-\left(\fr{\ta_2 \ta_3}{2(\ta_2 + \ta_3)}\right)(\partial_j B_j)^2 
\nonumber \\
&& 
\quad\quad
+\lambda (B_0^2 -B_i^2 - b^2) .
\label{bigH}
\eea
Four constraints are found for this model:
\bea
\phi_1& =& \Pi^{(\la)} ,
\label{c123lamphi1} 
\\
\phi_2 &=& - (B_0^2 -B_j^2 - b^2) ,
\label{c123lamphi2} 
\\ 
\phi_3 &=& - B_j \left[\fr{1}{\ta_1 - \ta_2}\Pi^j 
+\fr{\ta_1}{\ta_1 - \ta_2}(\partial_j B_0)\right] 
\nonumber \\
&& 
\quad\quad
+B_0\left[\fr{1}{\ta_2 + \ta_3}\Pi^0 
+ \fr{\ta_3}{\ta_2+\ta_3} (\partial_j B_j)\right] ,
\label{c123lamphi3} 
\eea
\bea
\phi_4 &=& 
-\lambda(B_0)^2
-\lambda \left(\fr{\ta_2 + \ta_3}{\ta_1 - \ta_2}\right)(B_j)^2
\nonumber \\
&& 
-\left(\fr{\ta_1 \ta_3}{2(\ta_1 - \ta_2)}
+\fr{\ta_1^2 (\ta_2 + \ta_3)}{2(\ta_1 - \ta_2)^2}\right)(\partial_j B_0)^2
\nonumber \\&& 
+\fr 1 2 \left(\fr{\ta_3^2}{\ta_2 + \ta_3}
+\fr{\ta_1 \ta_3}{\ta_1 - \ta_2}\right)(\partial_j B_j)^2
\nonumber \\
&& 
-\fr 1 2 (\ta_2 + \ta_3)B_j \partial_k \partial_k B_j
\nonumber \\&&
- \fr 1 2 \left(\fr{\ta_3^2 - (\ta_1 + \ta_3)
(\ta_2 + \ta_3))}{\ta_1 - \ta_2}\right)B_j \partial_j \partial_k B_k
\nonumber \\
&& 
+\left(\fr{\ta_1^2 - (\ta_1 - \ta_2)^2}{2(\ta_1 - \ta_2)}\right)
B_0 \partial_j \partial_j B_0
\nonumber \\&&
-\fr{\ta_3}{2(\ta_1 - \ta_2)}B_j (\partial_j \Pi^0) 
+\fr{\ta_1}{2(\ta_1 - \ta_2)}B_0 (\partial_j \Pi^j)
\nonumber \\
&& 
- \fr 1 {(\ta_1 - \ta_2)}\left(\fr 1 2 \ta_3
+\fr{\ta_1 (\ta_2 + \ta_3)}{\ta_1 - \ta_2}\right)\Pi^j \partial_j B_0
\nonumber \\&&
+\left(\fr{\ta_3}{\ta_1 + \ta_3}
+\fr{\ta_1}{2(\ta_1 - \ta_2)}\right)\Pi^0 (\partial_j B_j)
\nonumber \\
&& 
+\fr{1}{2(\ta_2 + \ta_3)}(\Pi^0)^2
-\fr{\ta_2 + \ta_3}{2(\ta_1 - \ta_2)^2}(\Pi^j)^2 .
\label{ta123lamphi4} 
\eea
The constraint $\ph_1$ is primary, 
while $\ph_2$, $\ph_3$, and $\ph_4$ are secondary.
All four are second-class.
According to Dirac's counting argument
there are $n - n_1 - n_2/2 = 5 - 0 - 4/2 = 3$ 
degrees of freedom in this model.

The constraint $\ph_2$ shows that only three of the four fields
$B_\mu$ are independent.
In the timelike case, it is natural to solve for $B_0$ in terms of $B_j$.
The first and third constraints can be used, respectively,
to fix $\Pi^{(\la)}$ to zero and to determine 
$\Pi^0$ in terms of $B_j$ and $\Pi^j$.
The remaining constraint $\ph_4$ can be used to determine $\la$
in terms of $B_j$ and $\Pi^j$.
Interestingly,
this leaves the same number of independent degrees of
freedom as in the KS bumblebee model with a similar potential.
One might have thought that switching from a Maxwell kinetic term,
which results in the removal of a primary constraint $\Pi^0 = 0$,
would have introduced an additional degree of freedom.
However,
instead, new secondary constraints appear that
still constrain $\Pi^0$,
though not to zero.
As a result,
$B_0$ and $\Pi^0$ remain unphysical 
degrees of freedom despite the change in the kinetic term.

Since the generalized bumblebee model is not
viewed as a modified theory of electromagnetism
(e.g., no current $J^\mu$ is introduced),
there is no analogue or modified version of Gauss' law 
as there is in the KS bumblebee model.
Nonetheless,
in the constraint $\ph_4$,
$\la$ plays a similar role as a nonlinear source term
for the other fields as it does in the KS bumblebee.
Indeed, the constraint equation $\ph_4 \approx 0$ reduces to the
same modified form of Gauss' law as in \rf{KSgauss1}
with $J^0 = 0$
in the limit where $\Pi^0 \rightarrow 0$
and the coefficients $\ta_1$, $\ta_2$, $\ta_3$
take Maxwell values.
Thus,
when considering initial values of the independent fields
$B_j$ and $\Pi^j$ in the generalized bumblebee case,
the constraint $\ph_4$ can play a role similar to that
of the modified Gauss's law in the KS bumblebee model.

Restrictions on the coefficients $\ta_1$, $\ta_2$, $\ta_3$
can be found by examining the freely propagating modes in the theory.
Investigations along these lines with gravity included
have been carried out by a number of authors 
\cite{tj08,c123}.
Since the theory with generalized kinetic terms
has three degrees of freedom,
there can be up to three independent propagating modes.
These include the NG modes associated with the
spontaneous Lorentz breaking.
To determine their behavior,
it suffices to work in a linearized limit and to
look for solutions in the form of harmonic waves.
Carrying this out in the Hamiltonian formulation
requires combining the linearized equations of motion 
to form a wave equation for $B_j$.
For physical propagation,
i.e., to avoid signs in the kinetic term that give rise to ghost modes,
the condition $(\ta_1 - \ta_2) > 0$ must hold
\cite{c123}.

In this case,
two solutions are found that propagate 
as transverse massless modes
at the speed of light.
However,
a third longitudinal mode can be found as well.
In an observer frame with wave vector $k_\mu = (k_0, 0, 0, k_3)$,
it obeys a zero-mass dispersion relation of the form
\beq
(\ta_1 - \ta_2) k_0^2 + (\ta_2 + \ta_3) k_3^2 = 0 .
\label{disp}
\eeq
For physical velocities,
the ratio 
\beq
\al \equiv {k_0^2} /{k_3^2} = -\fr {\ta_2 + \ta_3} {\ta_1 - \ta_2} 
\label{alpha}
\eeq
must be positive,
which together with the requirement of ghost-free propagation gives
\beq
(\ta_1 - \ta_2) > 0, \,\, (\ta_2 + \ta_3) < 0
\label{noghosts}
\eeq
Note in comparison that the KS bumblebee model has
$(\ta_2 + \ta_3) = 0$, 
and therefore the third degree of freedom does not 
propagate as a harmonic wave.
Instead, it is an auxiliary field that mainly
affects the static potentials
\cite{rbffak}.

The stability of the theory also depends on whether
${\cal H}$ is positive over the full phase space.
Examining this should include consideration of possible
initial values at $t=0$ that satisfy the constraints.
Using integration by parts and $\ph_2 \approx 0$,
the Hamiltonian \rf{bigH} can be written as the sum
of two parts,
\beq 
{\cal H} =  {\cal H}_\Pi + {\cal H}_B .
\label{Hsum}
\eeq
The first, 
\bea
{\cal H}_\Pi &=& 
\fr 1{2(\tau_1-\tau_2)} \left(\Pi^j+\tau_1\partial_jB_0\right)^2
\nonumber \\
&& 
+ \fr 1{2(\tau_2+\tau_3)} \left(\Pi^0+\tau_3\partial_jB_j\right)^2 ,
\label{HPi}
\eea
includes dependence on the momenta,
while the second,
\bea
{\cal H}_B &=& -{\tau_1-\tau_2\over2}(\partial_jB_0)^2
-{2\tau_1-\tau_2+\tau_3\over2}(\partial_jB_j)^2
\nonumber \\
&& 
\quad\quad
-{\tau_1-\tau_2\over4}(\partial_iB_j-\partial_jB_i)^2 ,
\label{HB}
\eea
depends only on the fields $B_\mu$.

First consider ${\cal H}_B$.
From the condition for ghost-free propagation in \rf{noghosts}, 
it follows that the first and third terms are nonpositive. 
The second term is nonpositive
as well if $2\tau_1-\tau_2+\tau_3>0$,
which implies $\al < 2$.
Thus ${\cal H}_B\le0$ if the conditions \rf{noghosts} hold
and $\al < 2$.

Next consider the momentum-dependent term ${\cal H}_\Pi$.
Assuming the conditions \rf{noghosts}
for ghost-free propagation, 
the first term is nonnegative, 
while the second is nonpositive. 
Note that the two terms are not independent, 
since they are related by constraint $\phi_3$.
However, one choice of initial values that makes 
both terms vanish
(and therefore satisfies $\ph_3 \approx 0$) is
\beq
\Pi^j+\tau_1\partial_jB_0=\Pi^0+\tau_3\partial_jB_j=0.
\label{5}
\eeq
The initial value of $\la$ is then chosen to make $\ph_4$ vanish,
and $B_0 = (b^2 + B_j^2)^{1/2}$ is used to make $\ph_2 \approx 0$.
Consequently, 
with ${\cal H}_\Pi$ vanishing, 
if $\alpha<2$, 
and the condition \rf{noghosts} holds,
then there exist initial conditions 
with ${\cal H}<0$.

To investigate the remaining cases, 
corresponding to other possible values of $\al$
consistent with \rf{noghosts},
use the constraint $\phi_3$ to rewrite
${\cal H}_\Pi$ as
\bea
{\cal H}_\Pi &=& {1\over2(\tau_1-\tau_2)}\left\{(\Pi^j+\tau_1\partial_jB_0)^2
\phantom{\fr B B}
\right.
\quad\quad
\nonumber \\
&& 
\left. \quad\quad
-\alpha\,
{[B_j(\Pi^j+\tau_1\partial_jB_0)]^2\over B_0^2}\right\}.
\label{6}
\eea
In any volume element, 
choose initial values for $B_j$ of the form 
$(B_1,B_2,B_3) = (0,0,B(\vec x))$.
It then follows that
\bea
{\cal H}_\Pi &=& {1\over2(\tau_1-\tau_2)}
\left\{(\Pi^1+\tau_1\partial_1B_0)^2+(\Pi^2+\tau_1\partial_2B_0)^2
\phantom{\fr B B}
\right.
\nonumber \\
&& 
\left.
+\left(1-\alpha{B^2\over b^2+B^2}\right)(\Pi^3+\tau_1\partial_3B_0)^2
\right\}.
\label{7}
\eea
With this form,
initial values of the components $\Pi^1$ and $\Pi^2$ 
can be chosen that make the first two
terms in this expression vanish. 
The third term becomes negative for any $\alpha>1$,
provided an initial value of $B^2$ is chosen that obeys
\beq
B^2>{b^2\over\alpha-1} .
\label{8}
\eeq
With ${\cal H}_\Pi < 0$,
and $\Pi^3+\tau_1\partial_3B_0\ne0$,
the initial value of $\Pi^3$ can then be made arbitrarily large
so that the total initial Hamiltonian density 
${\cal H}={\cal H}_\Pi + {\cal H}_B$ 
is negative, even if ${\cal H}_B>0$.

Thus, 
the Hamiltonian density ${\cal H}$ can take negative initial 
values for any choice of the parameters $\tau_1$, $\tau_2$, $\tau_3$ 
satisfying the conditions \rf{noghosts} for ghost-free propagation. 
The two examples with $\al < 2$ and $\al > 1$ are sufficient to
cover all possible cases.

Evidently a dilemma occurs in the generalized bumblebee model.
If the coefficients $\ta_1$, $\ta_2$, $\ta_3$ are restricted to
permit ghost-free propagation,
then regions of the full phase space allowed
by the constraints can occur with ${\cal H}<0$.
This parallels the behavior in the KS bumblebee model.
With $\ta_1$, $\ta_2$, $\ta_3$ equal to Maxwell values,
the allowed regions of phase space in the KS model include 
solutions with ${\cal H} < 0$.
However, as demonstrated in a previous section,
if initial values with $\la = 0$ are chosen,
and current conservation holds,
then $\la = 0$ and ${\cal H} > 0$ for all time
in the KS bumblebee model.

Based on this,
one could look for similar restrictions of the phase space
in the case of the generalized bumblebee model.
For example,
the solutions with ${\cal H} < 0$ described above
must typically have $\la \ne 0$ at $t=0$ 
to satisfy the constraint $\ph_4 \approx 0$.
This suggests the idea of trying to limit the choice of 
initial values to $\la = 0$ in an attempt 
to exclude the possibility of solutions with ${\cal H} < 0$ .

However,
this idea seems unlikely to succeed in the case of
the generalized bumblebee model,
since setting $\la = 0$ at $t=0$ is
not sufficient to restrict the phase space to solutions 
with $\la = 0$ for all time.
This is because the equation of motion for $\la$ has 
different dependence on the other fields in the generalized
bumblebee model compared to the KS model.
In particular,
$\dot \la$ is not proportional to just $\la$ itself.
This is evident even in the linearized theory,
with $B_\mu$ expanded as $B_\mu = b_\mu + \cE_\mu$.
Applying the constraint analysis to the linearized
theory yields a first-order expression for $\la$
in terms of $\cE_j$ and $\Pi^j$ equal to
\beq
\la \simeq \fr 1 {2b} \left( \fr {\ta_1 + \ta_3} { \ta_1 - \ta_2} \right) \partial_j \Pi^j ,
\label{laex2}
\eeq
while the equation of motion for $\la$ in the linearized theory is
\beq
\dot \la \simeq 
- \fr 1 {2b} \fr {(\ta_2 + \ta_3)(\ta_1 + \ta_3)} {(\ta_1 - \ta_2)}
(\partial_k \partial_k \partial_j \cE_j ) .
\label{ladot2}
\eeq
The latter equation shows that 
(with non-Maxwell values
$\ta_2 + \ta_3 \ne 0$)
$\dot \la$ is independent of $\la$ at linear order.
Therefore, even if $\la=0$ at $t=0$,
nonzero values of $\la$ can evolve over time.
This makes it difficult to decouple regions of
phase space with ${\cal H}>0$ in the generalized bumblebee model
purely by making a generic choice of initial values. 
It would thus seem likely that the regions of
phase space with ${\cal H} < 0$ include solutions
obeying $\la = 0$ at $t=0$.

\subsubsection{Quadratic Smooth Potential}

The generalized bumblebee model with a smooth quadratic
potential \rf{Vkappa} depends on four field components $B_\mu$
and their corresponding conjugate momenta.
The expressions for $\Pi^0$ and $\Pi^j$ are the same as
in Eqs.\ \rf{Pi0lam} and \rf{Pijlam}, respectively.
There are no constraints in this model.
Thus, according to Dirac's counting algorithm there
are $n - n_1 - n_2 = 4 - 0 - 0 = 4$ independent degrees
of freedom.
This is two more than in electromagnetism,
and one more than in the KS bumblebee model.

These four degrees of freedom include three NG modes
and a massive mode.
For arbitrary values of $\ta_1$, $\ta_2$, and $\ta_3$,
all three NG modes can propagate,
but with dispersion relations that depend on these coefficients.
In contrast, in the KS model,
with a Maxwell kinetic term,
only two of the NG modes propagate as transverse photons.
A massive mode occurs in either theory when
$V^\prime = 2 \ka (B_0^2 - B_j^2 - b^2) \ne 0$.
In the generalized bumblebee case,
there is no analogue of Gauss' law,
and it is possible for the massive mode to propagate.
However,
in the KS model with a timelike vector,
the constraint  \rf{KSsmoothPhi}  provides a
modified version of Gauss' law,
and the massive mode is purely an auxiliary field
that acts as a nonlinear source of charge density
in this relation.

The Hamiltonian for the generalized bumblebee
has the same form as in \rf{bigH},
but with the potential in the last term replaced by
the expression in \rf{Vkappa}.
With no constraints,
the full phase space includes solutions with an
unrestricted range of initial values.
Thus,
for any values of the coefficients $\ta_1$, $\ta_2$, $\ta_3$,
there will either be propagating ghost modes
or permissible initial choices for the 
fields and momenta with ${\cal H} < 0$.

\bigskip
\begin{table*}[t]
\caption{Summary of constraints.
Shown for each model are the number of
primary ($1^o$), secondary  ($2^o$), 
first-class (FC), and second-class (SC) constraints,
and the resulting number of independent degrees of 
freedom (DF).
The last column indicates the regions of phase space
that are ghost-free and have ${\cal H}>0$.
Current conservation $\partial_\mu J^\mu = 0$ is
assumed in the KS models.}
\begin{ruledtabular}
\begin{tabular}{|c|c|c|c|c|c|c|c|c|c|}
\small Theory & Kinetic Term 	& Potential $V$ 	& Fields 
& 1$^o$ & 2$^o$
& FC & SC 	& DF & Ghost-Free, ${\cal H}>0$
 \\ \hline
Electromagnetism		& $- \frac 1 4 F_{\mu\nu}F^{\mu\nu}$ & --
& $A_\mu$, $\Pi^\mu$  & 1	& 1
& 2 & 0	& 2	& full phase space
\\
 \hline
 Nambu Model		& $- \frac 1 4 F_{\mu\nu}F^{\mu\nu}$ & --
& $A_j$, $\Pi^j$  
& 0	& 0
& 0 & 0	& 3 & subspace ($\partial_j \Pi^j = J^0$)
	\\
 \hline
KS Bumblebee		
& $- \frac 1 4 B_{\mu\nu}B^{\mu\nu}$
& $ \lambda (B_\mu B^\mu \pm b^2)$ 	
& $B_\mu$, $\Pi^\mu$, $\lambda$, $\Pi^{(\la)}$  
& 2	& 2
& 0	& 4 	& 3 & subspace ($\la = 0$) 
\\
		& ($\ta_1 =1$, $\ta_2 = \ta_3 =  0$)
& $\frac 1 2 \kappa (B_\mu B^\mu \pm b^2)^2$ 
& $B_\mu$, $\Pi^\mu$   
& 1	& 1
& 0	& 2 	& 3 & subspace ($B_\mu B^\mu = b^2$) 
 \\ 
		& 
& $\frac 1 2 \lambda (B_\mu B^\mu \pm b^2)^2$ 
& $B_\mu$, $\Pi^\mu$, $\lambda$, $\Pi^{(\la)}$	
& 2	& 2
& 2	& 2 	& 2 & full phase space
\\ \hline
General Bumblebee		
& non-Maxwell
& $\lambda (B_\mu B^\mu \pm b^2)$ 	
& $B_\mu$, $\Pi^\mu$, $\lambda$, $\Pi^{(\la)}$  
& 1	& 3
& 0	& 4 	& 3 & no subspace found
 \\ 
		& (arbitrary $\ta_1$, $\ta_2$, $\ta_3$)
& $\frac 1 2 \kappa (B_\mu B^\mu \pm b^2)^2$ 	
& $B_\mu$, $\Pi^\mu$   
& 0	& 0	
& 0	& 0 	& 4 & no subspace found
 \\ 
		& 
& $\frac 1 2 \lambda (B_\mu B^\mu \pm b^2)^2$ 	
& $B_\mu$, $\Pi^\mu$, $\lambda$, $\Pi^{(\la)}$   
& 1	& 1
& 2	& 0 	& 3 & no subspace found
 \\
\end{tabular}
\end{ruledtabular}
\end{table*}

\subsubsection{Quadratic Lagrange-Multiplier Potential}

As a final example,
the generalized bumblebee model with a quadratic
Lagrange-multiplier potential \rf{Vsigma2} can be 
considered as well.
In this case there are ten fields $B_\mu$, $\Pi^\mu$,
$\la$, and $\Pi^{(\la)}$.
The conjugate momenta are given in 
\rf{Pi0lam} - \rf{Pilamlam}.
The Hamiltonian is the same as in \rf{bigH},
but with the potential replaced by \rf{Vsigma2}.
In this case,
two constraints are found,
\bea
\ph_1&=& \Pi^{(\la)} , \\
\ph_2 &=& - \fr 1 2 (B_0^2 - B_j^2 - b^2)^2 .
\label{lamquadconstraints}
\eea
Constraint $\ph_2$ imposes the condition \rf{condition}.
However, it involves a quadratic expression for this condition,
and therefore the system is irregular, 
and the same caveats must be applied as in the KS model.
In particular,
substitution of an equivalent constraint 
$\ph_2^\prime = (B_0^2 - B_j^2 - b^2)$ causes Dirac's
counting argument to fail.
However, with $\ph_1$ and $\ph_2$ identified as
first-class constraints,
Dirac's algorithm gives
$n - n_1 - n_2/2 = 5 - 2 - 0 = 3$ degrees of freedom.
This is again one more than in the KS model.
In this case,
there is no massive mode,
and $\la$ decouples completely.
The three independent degrees of freedom are the NG modes,
which in the generalized bumblebee can all propagate.
However, even if values of $\ta_1$, $\ta_2$, and $\ta_3$
can be found that prevent these modes from propagating
as ghost modes,
there are no other constraints in the theory that prevent initial-value
choices that can yield solutions with ${\cal H} < 0$.

\section{Summary \& Conclusions}

Table I summarizes the results of the constraint analysis applied
to electrodynamics, Nambu's model, the KS bumblebee, and
the generalized bumblebee.
For each of the bumblebee models,
three types of potentials $V$ are considered.
The results show that no two models have identical
constraint structures.
In most cases,
there are one or more additional degrees of freedom
in comparison to electromagnetism.
These extra degrees of freedom are important both
as possible additional propagating modes and in
terms of how they alter the initial-value problem.

In considering the stability of the bumblebee models,
it is not sufficient to look only at the propagating modes.
The range of possible initial values must be examined as well.
In general,
when the extra degrees of freedom 
appearing in these models are allowed access
to the full phase space,
the Hamiltonians are not strictly positive definite.
However,
in the KS models,
it is possible to choose initial values 
for the fields and momenta that
restrict the phase space to ghost-free regions with ${\cal H}>0$.
In contrast,
in models
with generalized kinetic terms
obeying $(\ta_2 + \ta_3) \ne 0$,
no such restrictions are found.
These theories either have propagating
ghosts or have extra degrees of freedom that evolve 
in such a way that makes it difficult to separate off
restricted regions of phase space with ${\cal H}>0$.
In the end,
it appears that only the KS models have a simple choice
of initial values that can yield a physically viable theory 
in a restricted region of phase space.

The examples considered in this analysis all focused
on the case of a timelike vector $B_\mu$,
which is the most widely studied case in the literature, 
since it involves an observer frame that maintains rotational invariance.
A natural extension of this work would 
be to consider models with a spacelike vector $B_\mu$.
In this case,
it is straightforward to show that the linearized KS model
is equivalent to electrodynamics in an axial gauge
\cite{rbak}.
However,
additional care is required in conducting
a constraint analysis of the full nonlinear KS or generalized models, 
since $B_0$ can vanish in the case of a spacelike vector,
making additional singularities a possibility.
Alternatively,
an analysis in terms of the BRST formalism could be pursued,
which would be suitable as well for addressing questions
of quantization.
Lastly,
an extension of the constraint analysis to a curved
spacetime in the presence of gravity would be relevant,
since ultimately
bumblebee models are of interest not only as effective field 
theories incorporating spontaneous Lorentz violation,
but also as modified theories of gravity.
For example,
they are currently one of the more widely used models for 
exploring implications of Lorentz violation 
in gravity and cosmology and in seeking alternative
explanations of dark matter and dark energy.
However, performing a constraint analysis with gravity
presents even greater challenges 
and is beyond the scope of this work.

In summary,
the constraint analysis presented here in a
flat-spacetime limit is useful in seeking
insights into the nature of theories with spontaneous
Lorentz violation and what their 
appropriate interpretations might be.
In particular,
the KS bumblebee models offer the possibility
that Einstein-Maxwell theory might emerge as a
result of spontaneous Lorentz breaking instead of
through local U(1) gauge invariance.
Indeed,
in the flat-spacetime limit of this model,
with a timelike vacuum value,
electromagnetism in a fixed nonlinear gauge
is found to emerge in a 
well-defined region of phase space.

\section*{Acknowledgments}

We thank Alan Kosteleck\'y for useful conversations.
This work was supported in part 
by NSF grant PHY-0554663.
The work of R. P.\ is supported by the 
Portuguese Funda{\c c}{\~a}o para a Ci{\^e}ncia e a Tecnologia.


\begin{widetext}

\begin{center}
Erratum Appended to Published Version
[Physical Review D {\bf 77}, 125007 (2008)]
\end{center}

\bigskip

The Hamiltonian density term ${\cal H}_B$
given in Eq.\ (60), in Section III D 1,
is incorrect.
A correct expression is
\begin{equation}
\quad\quad\quad\quad\quad\quad\quad\quad
\quad\quad\quad\quad
{\cal H}_B = \fr 1 2 (\tau_1-\tau_2)
\left[ (\partial_j B_k)^2 - (\partial_jB_0)^2 \right]
-\fr 1 2 (\tau_1+\tau_3)(\partial_jB_j)^2  .
\quad\quad\quad\quad\quad\quad\quad\quad 
\quad\quad\quad 
(60)
\nonumber
\end{equation}
This term is used to examine the positivity of the 
Hamiltonian density ${\cal H}$ for the case of a model
with a general kinetic term and a Lagrange-multiplier potential.
The change in the term ${\cal H}_B$ alters some of  
the conclusions that follow Eq.\ (60),
which are stated in terms of a parameter $\alpha$
defined in Eq.\ (56).
A revised argument still assumes $\alpha > 0$
and that Eq.\ (57) holds for ghost-free propagation.
The  argument and conclusions for $\alpha > 1$ are unchanged,
with the result that ${\cal H}$ can be negative.
However, for $0 < \alpha \le 1$,
a new examination has to be carried out.
Using Schwartz inequalities,
it is found that ${\cal H}_\Pi \ge 0$
and $(\partial_j B_k)^2 - (\partial_j B_0)^2 \ge 0$.
From the latter condition it follows that
${\cal H}_B \ge - \half (\tau_1 + \tau_3) (\partial_j B_j)^2$.
For $(\tau_1 + \tau_3) > 0$,
corresponding to $\alpha <1$,
${\cal H}_B$ can be made arbitrarily negative,
and solutions with ${\cal H} < 0$ can therefore exist.
However, for $(\tau_1 + \tau_3) = 0$,
corresponding to $\alpha =1$,
it follows that ${\cal H}_B \ge 0$,
and that the Hamiltonian density ${\cal H}$ is nonnegative.
The model with $\alpha = 1$ and ${\cal H} \ge 0$ has
a Lagrangian density with a kinetic term proportional to
${\cal L} = -\half (\partial_\mu B_\nu)(\partial^\mu B^\nu)$.
It has recently been examined in arXiv:0812.1049
by S.M.\ Carroll, T.R.\ Dulaney, M.I.\ Gresham, and H. Tam.
It provides a counterexample to the conclusions
summarized in Table I for the
case with general values of $\tau_1$, $\tau_2$, $\tau_3$
and a linear Lagrange-multiplier potential.
The model with the same kinetic term 
but with a quadratic Lagrange-multiplier 
(as discussed in Section III D 3) 
also has ${\cal H}>0$ when $(\tau_1 + \tau_3) = 0$.
However,
all of the other results in the paper,
including the discussion of the KS model and
the smooth quadratic potential, 
are unchanged by this correction.

\end{widetext}


\begin{thebibliography}{99}
\bibitem{BBnote}
The inspiration for this name (given by Kosteleck\'y)
comes from the real-life bumblebees.
For a long time it was thought that with their large
bodies and little wings 
it was theoretically impossible for them to fly,
and yet they do.
Similarly, in bumblebee models the usual
theoretical explanation for massless photons,
local U(1) gauge invariance,
does not apply.
Nonetheless, massless NG modes do propagate
(as photons in the KS models)
due to the spontaneous Lorentz 
violation in these models.
\end{thebibliography}
\end{document}